\documentclass[5p]{elsarticle}
 \usepackage{lipsum}

\graphicspath{{figures/}}

\usepackage{lineno}
\modulolinenumbers[5]
\usepackage{epstopdf}%This line makes .eps figures into .pdf - please comment out if not required.
\usepackage{subcaption}
\usepackage{tabularx}
\usepackage{graphics} 
\usepackage{graphicx} % Allows including images
\usepackage{caption}
\usepackage{subcaption}
\usepackage{amsmath,amssymb,amstext,amsthm,mathtools}
\usepackage{float}
\usepackage{booktabs}
\usepackage{multirow}
\usepackage{siunitx} 
\usepackage{array,bm,color,epsfig}
\usepackage{caption}
\usepackage{url}
\usepackage[english]{babel}
\usepackage[utf8x]{inputenc}
\usepackage{setspace}
\usepackage{filecontents}
\usepackage{natbib}
\usepackage{float}
\usepackage{listings}
\usepackage{color}
\usepackage{xcolor}
\usepackage{wasysym}
\usepackage[flushleft]{threeparttable}
\usepackage{hhline}
\usepackage{multicol,tabularx,capt-of,ragged2e,ar}
\usepackage{mathrsfs}
\usepackage{tikz}
\usepackage{pifont}
\usepackage{stmaryrd}
\usepackage{pdflscape}
\usepackage{mathrsfs}
\usepackage{stmaryrd}
\usepackage{fancyheadings}
\usepackage{wrapfig,lipsum,booktabs}
\usepackage{soul}
\usepackage[toc]{appendix}
\usepackage{footnote}
\def\dt{\partial t}

\def\BE{\begin{equation}}
\def\EE{\end{equation}}

\usepackage[colorlinks]{hyperref}

\usepackage{scalerel,stackengine}
\stackMath
\newcommand\reallywidehat[1]{%
\savestack{\tmpbox}{\stretchto{%
  \scaleto{%
    \scalerel*[\widthof{\ensuremath{#1}}]{\kern.1pt\mathchar"0362\kern.1pt}%
    {\rule{0ex}{\textheight}}%WIDTH-LIMITED CIRCUMFLEX
  }{\textheight}% 
}{2.4ex}}%
\stackon[-6.9pt]{#1}{\tmpbox}%
}
\journal{}

\begin{document}

\begin{frontmatter}

\title{A Parallel Integrated Computational-Statistical Platform \\ for Turbulent Transport Phenomena}

\author[ME,CMSE]{Ali Akhavan-Safaei}
\author[ME,STAT]{Mohsen Zayernouri\corref{corr_author}}

\cortext[corr_author]{Corresponding author}\ead{zayern@msu.edu}

\address[ME]{Department of Mechanical Engineering, Michigan State University, East Lansing, MI 48824, USA}
\address[CMSE]{Department of Computational Mathematics, Science and Engineering, Michigan State University, East Lansing, MI 48824, USA}
\address[STAT]{Department of Statistics and Probability, Michigan State University, East Lansing, MI 48824, USA}

\begin{abstract}
In this paper, we present an open-source, automated, and multi-faceted computational-statistical platform to obtain synthetic homogeneous isotropic turbulent flow and passive scalar transport. A parallel implementation of the well-known pseudo-spectral method in addition to the comprehensive record of the statistical and small-scale quantities of the turbulent transport are offered for executing on distributed memory CPU-based supercomputers. The user-friendly workflow and easy-to-run design of the developed package is disclosed through an extensive and step-by-step example. The resulting low- and high-order statistical records vividly verify well-established and fully-developed turbulent state as well as the seamless statistical balance of conservation laws. Post-processing tools provided in this platform would let the user to readily construct multiple important transport quantities from the primitive turbulent fields.

\end{abstract}

\begin{keyword}
Turbulent Transport \sep Passive Scalars \sep  Statistical Analysis \sep Pseudo-spectral Method \sep High-Performance Computing
\end{keyword}

\end{frontmatter}

% \linenumbers

\begin{table}[t!]
	\centering
	    \caption*{Code Metadata}
	    \vspace{-.1 in}
	    \centering
	    {\scriptsize
    	\begin{tabular}{c c}
    		\toprule \toprule
    		Version &  v1.0\\
    		Link to code/repository &  \href{https://github.com/FMATH-Group/PSc_HIT3D}{\textcolor{black}{\texttt{github.com/FMATH-Group/PSc\_HIT3D}}}\\
    		Legal Code Licence & General Public License (GPLv3)\\
    		Code versioning system used &  git\\
    		Software code languages &  Python, MPI\\
    		Requirements \& dependencies  & NumPy, SciPy, MPI4Py\\
    		Support email for questions  &  \href{mailto: akhavans@msu.edu}{\textcolor{black}{\texttt{akhavans@msu.edu}}}\\
    		\bottomrule \bottomrule
    	\end{tabular}
    	}
\end{table}

% =============================================================================
%                           Motivation & Significance
% =============================================================================
\section{Motivation and Significance}\label{sec: Intro}
% Understanding the complex and random nature of turbulent flows, mixing, and transport is a vital step in predictions and the design of systems interacting with such heterogeneous medium. Turbulence is inherently consisted of multi-scale processes that requires high-accurate measurements at the smallest scales of transport. Direct numerical simulation (DNS) is a widely popular and extremely useful research tool that exploits powerful numerical schemes and computational techniques to solve the proper system of partial differential equations (PDEs) associated with conservation laws that constitute the dynamics of transport. 
Developing an open-source, sustainable, portably parallel, and integrated computational-statistical platform with high-order spatial and temporal accuracy provides a useful academic ground for better understanding of complex standard to anomalous turbulent transport across a multitude of scales \citep{akhavan2020anomalous}. Moreover, from the educational point of view, developing such a user-friendly scientific software will essentially fill the existing training gap in the subjective trinity, \textit{i.e.}, fluid mechanics, computational fluid dynamics, and turbulent transport; hence, leading to a more cohesive ramp up in training the future generation of researchers in a variety of academic-to-industrial disciplines.

Direct numerical simulation (DNS) of turbulent transport as a rigorous scientific tool is supposed to fully resolve the smallest scales of the motion resulted from the fluctuating fields in spatial domain while maintaining a high-order temporal accuracy as turbulence evolves in time \citep{moin1998DNS}. Among the current open-source computational platforms, \texttt{Nektar++} \citep{nektar++-2015, nektar++-2020} (the spectral/$hp$ element method flow software), \texttt{HERCULIS} \citep{herculis-2016} and \texttt{Xcompact3D} \citep{xcompact3d-2020} (the high-order finite difference flow solvers), \texttt{GRINS} \citep{grins2016} (the adaptive mesh refinement finite element method software), \texttt{spectralDNS} \citep{MORTENSEN2016} (the spectral method computational package for DNS), and \texttt{OpenFOAM} \citep{OpenFOAM}  are the notable contributions to the DNS of turbulent transport. On the other hand, the random nature of turbulence requires a thorough statistical analysis on the fluctuating fields and their gradients so that one can identify when the realistic and fully-developed turbulent state is obtained in DNS during an ongoing simulation. This necessitates development of a comprehensive computational platform that includes computing and recording of such statistical quantities of turbulent transport as time-series format.
% Given such fact, performing an accurate DNS is a computationally demanding task that urges the developers to leverage a wide spectrum of modern computational proficiency where the state of the art high-performance computing (HPC) techniques are their essential backbone.

\begin{figure*}[t!]
    \begin{minipage}[b]{.49\linewidth}
        \centering
        \includegraphics[width=.85\textwidth]{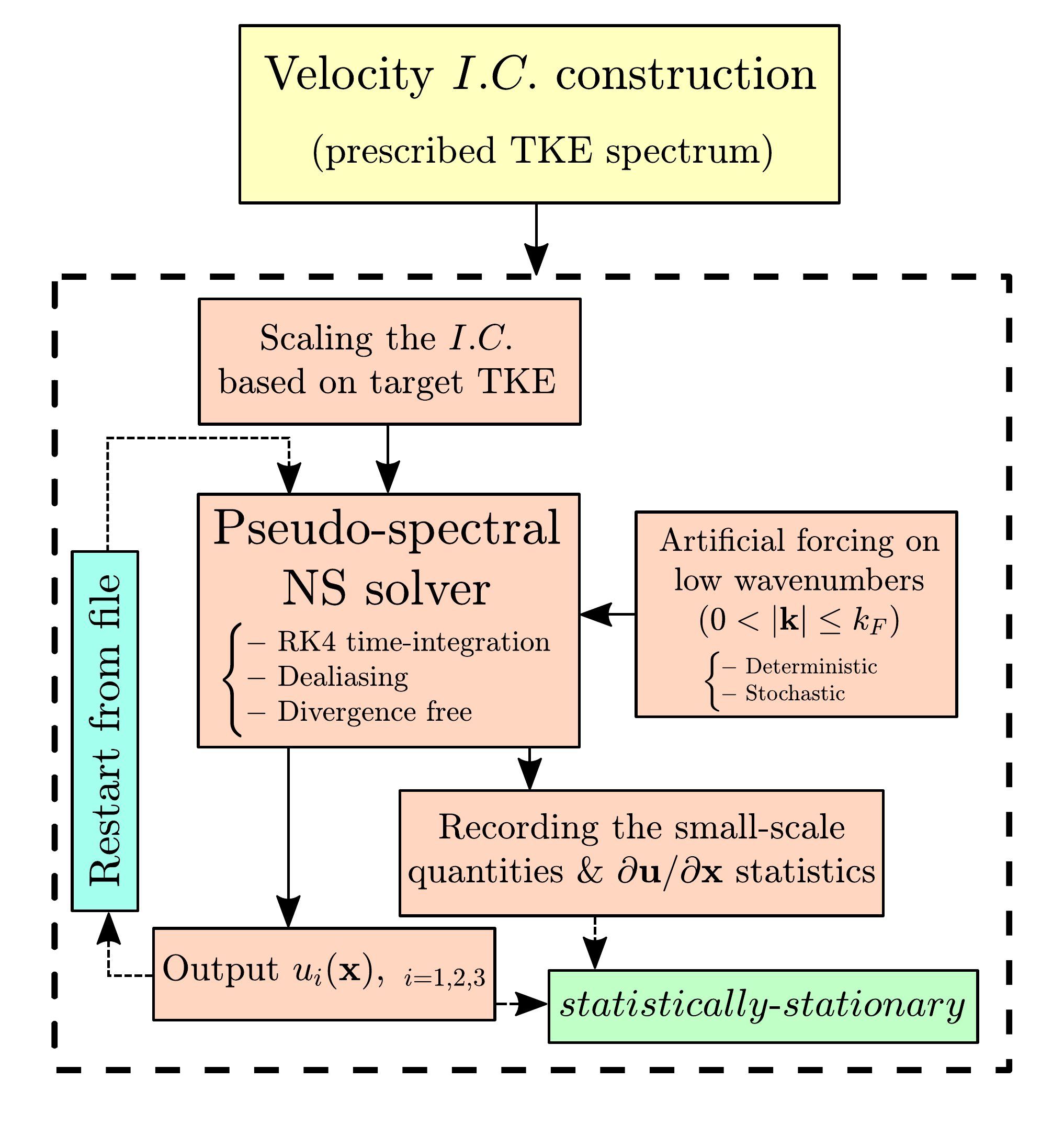}
        \subcaption{}\label{fig: Schematic_HIT}
    \end{minipage}
    \begin{minipage}[b]{.49\linewidth}
        \centering
        \includegraphics[width=.85\textwidth]{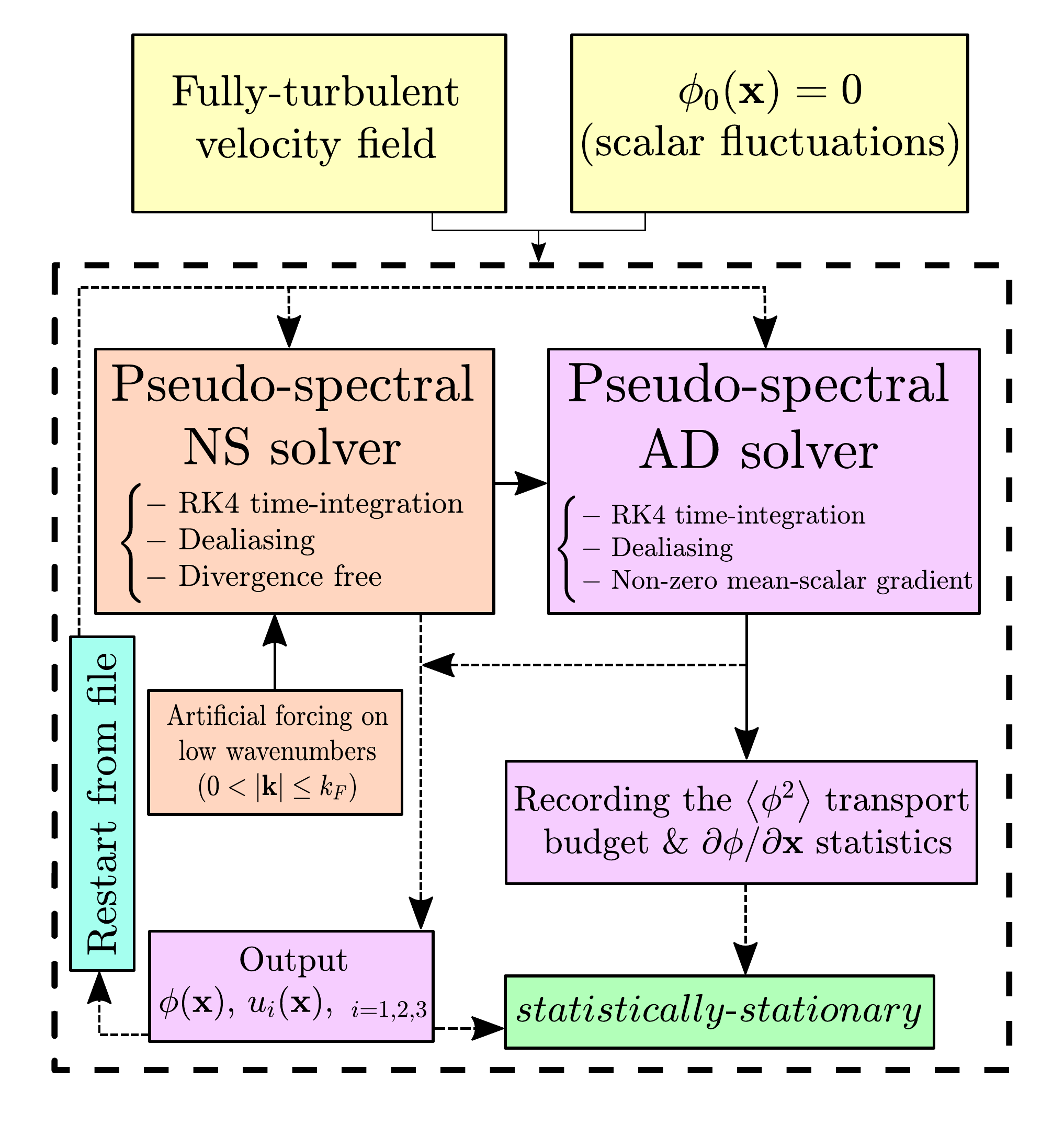}
        \subcaption{}\label{fig: Schematic_Scalar}
    \end{minipage}
    \caption{Schematic of the architecture of the software. Fig. (a) Illustrates the pseudo-spectral NS solver to archive fully-developed turbulent state, statistical records, velocity output and restarting the simulation from output file. Fig. (b) Shows the pseudo-spectral NS and AD solvers to reach fully-developed turbulent scalar state. }\label{fig: Schematics}
\end{figure*}

In the current work, our goal is to offer an extensible open-source computational platform that carries out the high-fidelity simulations of homogeneous isotropic turbulence (HIT) for an incompressible flow and also obtains the transport of a passive scalar (temperature or concentration of species) in such flow \citep{warhaft2000passive, shraiman2000scalar} while it keeps track of statistical quantities of turbulent transport. Here, we numerically solve the incompressible Navier-Stokes (NS) equations in addition to the advection-diffusion (AD) to sufficiently resolve the fluid velocity and passive scalar fields, respectively. Spatial homogeneity of fluctuating fields makes this problem well-suited for pseudo-spectral implementation of the NS and AD equations based on the Fourier collocation discretization as employed in our work. This computational platform is based upon programming in \texttt{PYTHON} and leveraging MPI library for parallel implementation.

The rest of this paper is organized as follows: in section \ref{sec: Software}, we describe the details and capabilities of the developed platform as a scientific software and we point out the theoretical backgrounds briefly. Furthermore, in section \ref{sec: Examples}, we go over a comprehensive example illustrating the results of a fully-developed turbulent flow and passive scalar field with proper statistical testing and verification. In section \ref{sec: Impact}, we outline broader impacts of the current work onto the research in turbulent transport. Finally, in section \ref{sec: Conclusion}, we conclude this paper with a summary and conclusions. 

% =============================================================================
%                               Software Description
% =============================================================================
\section{Software Description}\label{sec: Software}
\subsection{Governing Equations}\label{sec: Gov-Eqns}
The incompressible HIT considered in the present software is governed by the NS equations
\begin{align}\label{eqn: NS}
	\frac{\partial \boldsymbol{U}}{\dt} + \boldsymbol{U} \cdot \nabla \boldsymbol{U} &= -\nabla p + \nu \, \Delta \boldsymbol{U} + \mathcal{A} \, \boldsymbol{U},
\end{align}
subject to the continuity $\nabla \cdot \boldsymbol{U}=0$. In \eqref{eqn: NS}, $\boldsymbol{U}=(U_1,U_2,U_3)$ and $p$ are the instantaneous velocity and modified pressure (pressure divided by the constant density of fluid) fields in the Cartesian coordinate system $\boldsymbol{x}=(x_1,x_2,x_3)$, respectively. Moreover, $\nu$ denotes the dynamic viscosity of the Newtonian fluid, and $\mathcal{A}$ is a dynamically evaluated coefficient corresponding to the artificial forcing scheme we employ in order to obtain statistically stationary and fully-turbulent state.
From the Reynolds decomposition of instantaneous velocity field, $\boldsymbol{U}(\boldsymbol{x},t) = \left\langle \boldsymbol{U}(\boldsymbol{x},t) \right\rangle + \boldsymbol{u}(\boldsymbol{x},t)$, where $\langle \cdot \rangle$ represents the ensemble-averaging operator, and $\boldsymbol{u}(\boldsymbol{x},t)$ denotes the fluctuating part of the velocity field. In HIT, $\langle \boldsymbol{U}(\boldsymbol{x},t) \rangle=0$; therefore, the instantaneous velocity field equals the fluctuating part that is governed by \eqref{eqn: NS}. Introducing a passive scalar $\Phi(\boldsymbol{x},t)$ transported in the considered fully-developed HIT flow, the AD equation governing the passive scalar concentration may be formulated as
\begin{align}\label{eqn: AD-total}
	\frac{\partial \Phi}{\dt} + \boldsymbol{u} \cdot \nabla \Phi &= \mathcal{D} \, \Delta \Phi,
\end{align}
where $\mathcal{D}$ denotes the diffusivity of passive scalar. Applying the Reynolds decomposition on the total passive scalar, $\Phi = \langle \Phi \rangle + \phi$, and $\phi$ is the fluctuating part of the scalar concentration. Considering a uniform mean gradient for the passive scalar as $\nabla \langle \Phi \rangle=(0,\beta,0)$, where $\beta$ is a constant, the AD equation in \eqref{eqn: AD-total} is rewritten as
\begin{align}\label{eqn: AD}
	\frac{\partial \phi}{\dt} + \boldsymbol{u} \cdot \nabla \phi &= -\beta \, u_2 + \mathcal{D} \, \Delta \phi.
\end{align}

\subsection{Fourier Pseudo-Spectral Method}\label{sec: Pseudo-Spectral}

Here, we consider spatial homogeneity for the fluctuating velocity and scalar concentration, which allows periodic boundary conditions for these fluctuating fields as 
\begin{align}\label{eqn: BC}
	\boldsymbol{u}(\boldsymbol{x}+\mathcal{L} \, \mathrm{e}_i,t) = \boldsymbol{u}(\boldsymbol{x},t), \quad \phi(\boldsymbol{x}+\mathcal{L} \, \mathrm{e}_i,t) = \phi(\boldsymbol{x},t),
\end{align}
where $\mathrm{e}_i, \: _{i=1,2,3}$, is the unit vector for the $i$-th direction of the Cartesian coordinate, and $\mathcal{L}$ is the periodicity length that defines the spatial domain as $\boldsymbol{\Omega} = [0,\mathcal{L}]^3$. Discretizing $\boldsymbol{\Omega}$ using a uniform three-dimensional grid returns $N^3$ grid points with grid spacing along each direction as $\Delta x = \mathcal{L}/N$. Transforming this discretization into spectral domain let us have a standard pseudo-spectral representation of the governing equations \eqref{eqn: NS} and \eqref{eqn: AD}. Subsequently, $\boldsymbol{k}=(k_1,k_2,k_3)$ represents the coordinate system in the spectral space and using Fourier collocation method the discretized representation of $\boldsymbol{k}$ would be $k_i=(-N/2+1,\dots,N/2)$, $i=1,2,3$. Accordingly, the discrete Fourier transform of any field variable such as $\phi(\boldsymbol{x},t)$ is written as
\begin{align}\label{eqn: DFT}
    \phi(\boldsymbol{x},t) = \frac{1}{N^3} \sum_{\boldsymbol{k}} \hat{\phi}_{\boldsymbol{k}}(t) \, e^{\mathfrak{i}\boldsymbol{k}\cdot \boldsymbol{x}},
\end{align}
where $\mathfrak{i}=\sqrt{-1}$, and $e^{\mathfrak{i}\boldsymbol{k}\cdot \boldsymbol{x}}$ are the Fourier basis functions. Subsequently, the Fourier coefficients associated with $\boldsymbol{k}$ are represented as $\hat{\phi}_{\boldsymbol{k}}(t) = \sum_{\boldsymbol{x}} \phi(\boldsymbol{x},t) \, e^{-\mathfrak{i}\boldsymbol{k} \cdot \boldsymbol{x}}$. Standard pseudo-spectral formulation of the NS equations based upon the Fourier collocation method is obtained after taking the Fourier transform of \eqref{eqn: NS},
\begin{align}\label{eqn: P-S_NS1}
	\frac{d\hat{\boldsymbol{u}}_{\boldsymbol{k}}}{dt} + (\reallywidehat{\boldsymbol{u} \cdot \nabla \boldsymbol{u}})_{\boldsymbol{k}} &= -\mathfrak{i}\boldsymbol{k} \: \hat{p}_{\boldsymbol{k}} -\nu \vert \boldsymbol{k} \vert^2 \hat{\boldsymbol{u}}_{\boldsymbol{k}} + \mathcal{A} \hat{\boldsymbol{u}}_{\boldsymbol{k}}; \\ 
	\mathfrak{i}\boldsymbol{k} \cdot \hat{\boldsymbol{u}}_{\boldsymbol{k}}=0, \nonumber
\end{align}
By taking the divergence of momentum equation in \eqref{eqn: P-S_NS1} and applying the continuity, modified pressure is explicitly represented in terms of the velocity field. Considering that  ${\boldsymbol{k}} \cdot {\boldsymbol{k}} = \vert \boldsymbol{k} \vert^2$, one can derive $\hat{p}_{\boldsymbol{k}}=\mathfrak{i} \boldsymbol{k}\cdot (\reallywidehat{\boldsymbol{u} \cdot \nabla \boldsymbol{u}})_{\boldsymbol{k}}/\vert \boldsymbol{k} \vert^2$; hence, \eqref{eqn: P-S_NS1} may be reformulated as
\begin{align}\label{eqn: P-S_NS2}
	\frac{d\hat{\boldsymbol{u}}_{\boldsymbol{k}}}{dt} + (\reallywidehat{\boldsymbol{u} \cdot \nabla \boldsymbol{u}})_{\boldsymbol{k}} &= \boldsymbol{k} \frac{\boldsymbol{k}\cdot (\reallywidehat{\boldsymbol{u} \cdot \nabla \boldsymbol{u}})_{\boldsymbol{k}}}{\vert \boldsymbol{k} \vert^2} -\nu \vert \boldsymbol{k} \vert^2 \hat{\boldsymbol{u}}_{\boldsymbol{k}} + \mathcal{A} \hat{\boldsymbol{u}}_{\boldsymbol{k}}.
\end{align}
Similarly, the pseudo-spectral representation of the AD equation for passive scalar \eqref{eqn: AD}, is written as
\begin{align}\label{eqn: P-S_AD}
	\frac{d\hat{\phi}_{\boldsymbol{k}}}{dt} + (\reallywidehat{\boldsymbol{u} \cdot \nabla \phi})_{\boldsymbol{k}} &= -\beta \, \hat{u}_{k_2} -\mathcal{D} \vert \boldsymbol{k} \vert^2 \hat{\phi}_{\boldsymbol{k}}.
\end{align}
Employing the fourth-order Runge-Kutta (RK4) scheme, the time-stepping for both NS and AD equations is explicitly done since the nonlinear (advetctive) terms are evaluated in the physical space and then transformed into the spectral space.

\begin{table*}[t!]
	\centering
	    \caption{Statistical characteristics of turbulent flow to be recorded from the NS solver as time-series within user-defied time-intervals.}\label{tab: small-scale_chr}
	    \centering
    	\begin{tabular}{c c c c c}
    		\toprule \toprule%hline \hline \\
    		TKE &
    		Dissipation &
    		Kolmogorov length-scale &
    		Taylor-scale Reynolds &
    		Large-eddy turnover time \\ \midrule 
    		$K = \frac{1}{2}\langle \boldsymbol{u} \cdot \boldsymbol{u} \rangle$ &
    		$\varepsilon = 2 \, \nu \, \langle \boldsymbol{S}:\boldsymbol{S} \rangle$ &
    		$\eta = \left( \nu^3/\varepsilon \right)^{1/4}$ &
    		$Re_\lambda = \lambda \, u_{rms}/\nu$ &
    		$T_e = l_o/u_{rms}$
    		\vspace{0.1 in} \\
    		\bottomrule \bottomrule%\hline \hline
    	\end{tabular}
    	\begin{center}
    	    \begin{tablenotes}
                \item[$\ast$] {\footnotesize $\boldsymbol{S}=\frac{1}{2} \left(\nabla \boldsymbol{u}+\nabla \boldsymbol{u}^\mathrm{T} \right), \qquad u_{rms}=\sqrt{2K/3}, \qquad \lambda = u_{rms}\sqrt{15 \nu / \varepsilon}, \qquad l_o = u_{rms}^3/\varepsilon$.}
            \end{tablenotes}
    	\end{center}
	   % \vspace{0.1 in}
\end{table*}

% =============================================================================
%                           Software Architecture
% =============================================================================
\subsection{Software Architecture}\label{sec: Architecture}

\begin{figure*}[t!]
    \centering
    \begin{minipage}[b]{.39\linewidth}
        \centering
        \includegraphics[width=.85\textwidth]{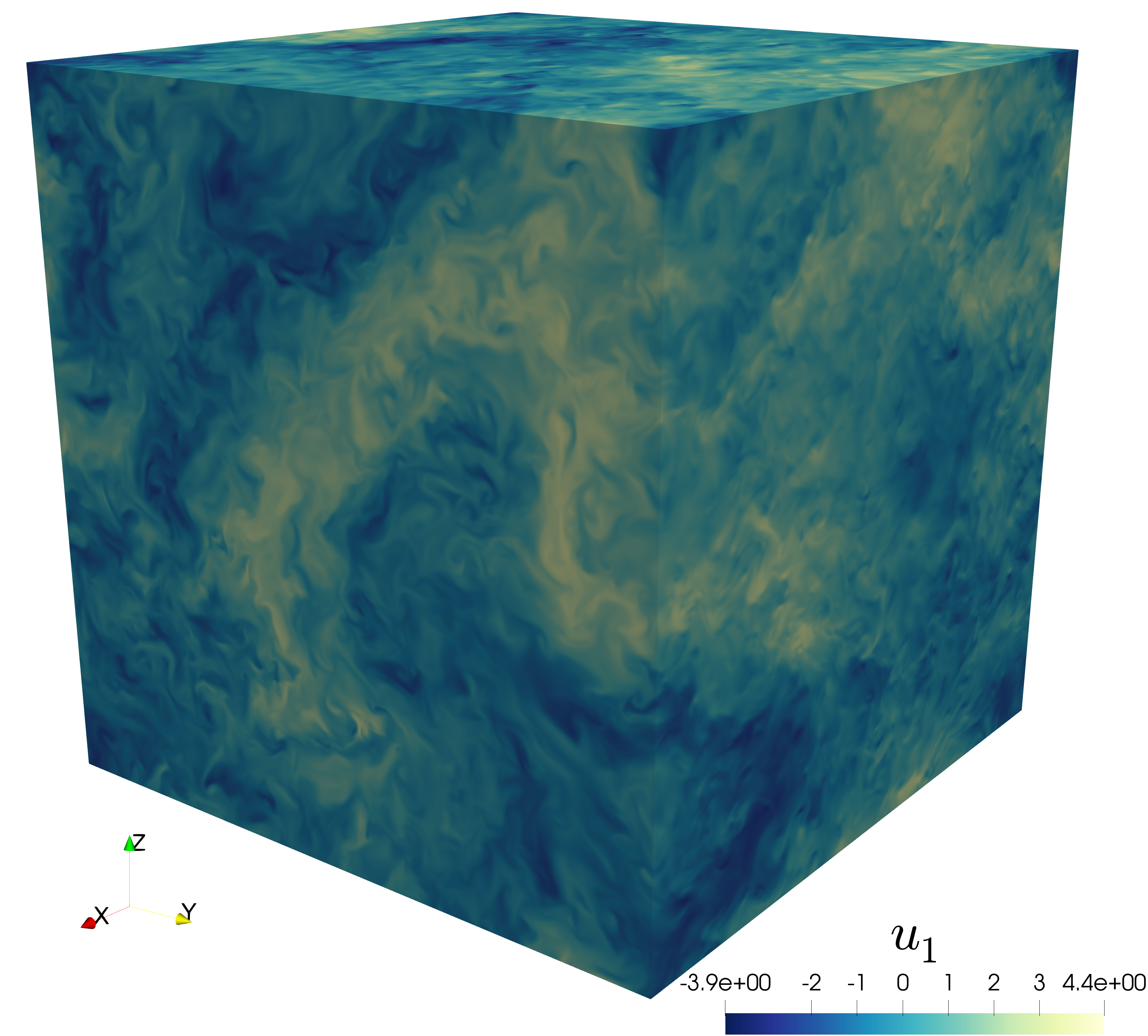}
        \subcaption{}\label{fig: HIT_u1}
    \end{minipage}
    \begin{minipage}[b]{.01\linewidth}
        ~
    \end{minipage}
    \begin{minipage}[b]{.56\linewidth}
        \centering
        \includegraphics[width=.85\textwidth]{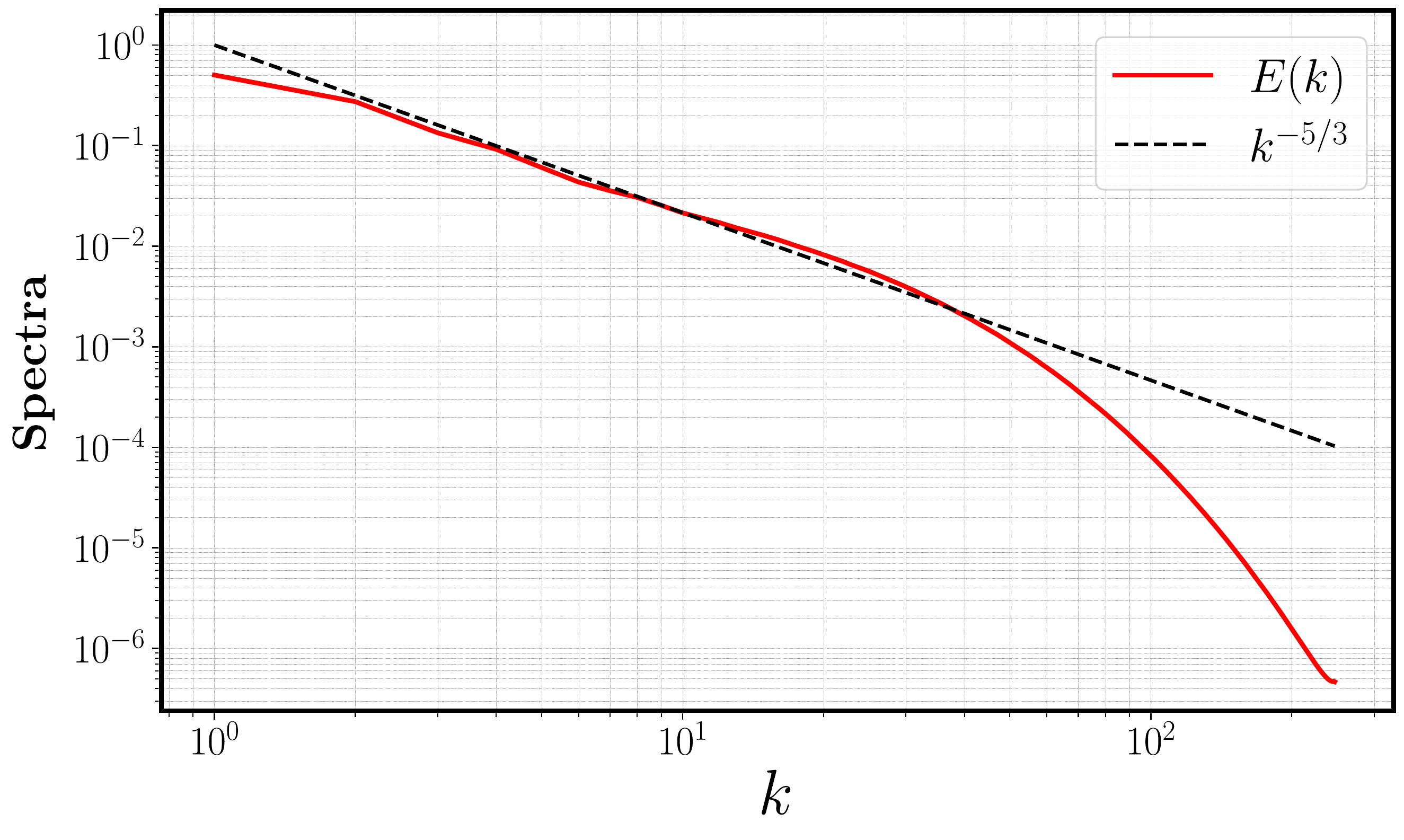}
        \subcaption{}\label{fig: TKE_spectrum}
    \end{minipage}
    \begin{minipage}[b]{.65\linewidth}
        \centering
        \includegraphics[width=1\textwidth]{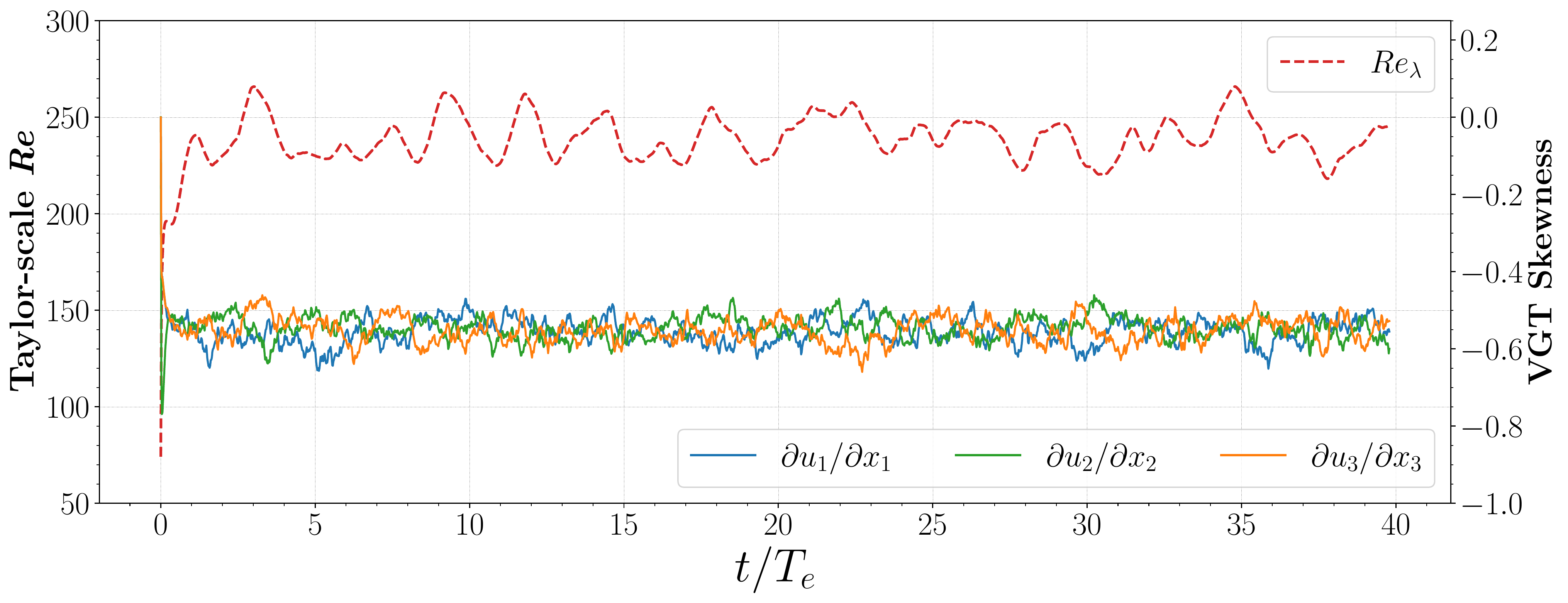}
        \subcaption{}\label{fig: HIT_records}
    \end{minipage}
    \caption{(a) Snapshot of fully-developed turbulent velocity field, $u_1$ component. (b) Time-averaged TKE spectrum. (c) Time-series of $Re_\lambda$ (red dashed line), and VGT skewness factors, $\mathcal{S}_{u_{1,1}}$, $\mathcal{S}_{u_{2,2}}$, $\mathcal{S}_{u_{3,3}}$ .}\label{fig: HIT_velocity}
\end{figure*}

The structure of the prepared software is schematically illustrated in Figure \ref{fig: Schematics}. According to Figure \ref{fig: Schematic_HIT}, a user starts from a pre-processing step, where the isotropic and random velocity initial condition (IC) is constructed based on a prescribed spectrum for turbulent kinetic energy (TKE). The procedure is the straightforward implementation of the well-known work by Rogallo to generate divergence-free isotropic velocity state \citep{Rogallo1981numerical}. According to Lamorgese \textit{et al.} \citep{Lamorgese2005}, the initial TKE spectrum is chosen to be
\begin{align}\label{eqn: TKE_Spectrum_IC}
	E(\kappa,0) =\frac{u_{rms}^2}{k_F} \times\begin{cases}
			    (\kappa/k_F)^{2}, & \text{if} \ \ \ \kappa \leq k_F,\\
			    (\kappa/k_F)^{-5/3}, & \text{if} \ \ \ \kappa > k_F.
			  \end{cases}
\end{align}
where $\kappa$ represents the wavenumber associated with spherical shells, $k_F$ denotes the maximum  wavenumber of TKE shell we apply artificial forcing to, and $u_{rms}$ specifies the initial root-mean-square (rms) intensity of velocity fluctuations. In construction of velocity IC, $u_{rms}$ is set to be unity while $k_F$ and the number of Fourier collocation points, $N$ are taken as input parameters. In a \texttt{UNIX/LINUX} environment, these inputs are taken as arguments in the execution commandline that are imported through \texttt{sys} library in \texttt{PYTHON}. Once the velocity IC is obtained, it is partitioned into $N_p$ slabs according to the slab decomposition method. We completely adopted the implementation of Mortensen and Langtangen \citep{MORTENSEN2016} for domain decomposition in addition to the parallel implementation of forward and inverse three-dimensional fast Fourier transform (FFT) in \texttt{PYTHON} programming language. Here, the MPI communications depend on the \texttt{mpi4py} library \citep{dalcin2005mpi, dalcin2008mpi, dalcin2011parallel}.

Having the partitioned velocity IC prepared in the pre-processing step, it is fed to the main body of the software where the initial velocity field might be magnified by a user-defined input argument so that a target TKE is considered for the simulation. The viscosity of fluid, $\nu$, is also is taken as another user-defined input argument. Next, the magnified velocity field is passed into the solver where $\hat{\boldsymbol{u}}_{\boldsymbol{k}}$ and $(\reallywidehat{\nabla \boldsymbol{u}})_{\boldsymbol{k}}=\mathfrak{i}\boldsymbol{k}\hat{\boldsymbol{u}}_{\boldsymbol{k}}$ are separately transformed back into the physical space so that $\boldsymbol{u} \cdot \nabla \boldsymbol{u}$ is simply computed and then transformed into Fourier space. The aliasing error that appears due to this procedure is removed by phase-shifting and truncation according to $2\sqrt{2}N/3$ as the maximum wavenumber \citep{de-Aliasing_1971}. Afterwards, all of the terms in equation \eqref{eqn: P-S_NS2} are directly evaluated in every stage of RK4 time-integration; however, the last term in the right-hand side of \eqref{eqn: P-S_NS2} is only evaluated after the last stage during the artificial forcing by keeping the energy of the low wavenumbers constant, which is associated with the sphere of $0 < \vert \boldsymbol{k} \vert \leq k_F$. In this procedure, $\mathcal{A}$ is computed in a way that the the dissipated energy of turbulent motion is injected to the large-scales. This scheme prevents the flow to undergo decay process before the realistic and fully-developed turbulent state is achieved, nevertheless, the artificial forcing scheme could be turned off through a user-defined input argument if one seek to obtain decaying HIT data. The forcing coefficient could be determined either deterministically \cite{Sullivan1994_deterministic-forcing} or stochastically \citep{ESWARAN1988, Alvelius1999_random-forcing} and both of these methods are supported in the software as would be specified as an input option. Moreover, regarding the stable time-integration, the Courant-Friedrichs-Lewy (CFL) number is dynamically checked through a user-defined time-frequency. According to Eswaran and Pope's work \citep{ESWARAN1988}, CFL for this problem is demonstrated as
\begin{align}\label{eqn: CFL}
    \mathrm{CFL} := \frac{\Delta t}{\Delta x}\max \left( \vert u_1 \vert + \vert u_2 \vert + \vert u_3 \vert \right).
\end{align}
In \eqref{eqn: CFL}, $\Delta x$ is the uniform grid spacing in each direction and $\Delta t$ is the user-defined constant time-interval used in the RK4 time-stepping. In practice, CFL is required to be less than unity to ensure a stable time-integration.

\begin{figure*}
    \begin{minipage}[b]{.64\linewidth}
        \centering
        \includegraphics[width=1\textwidth]{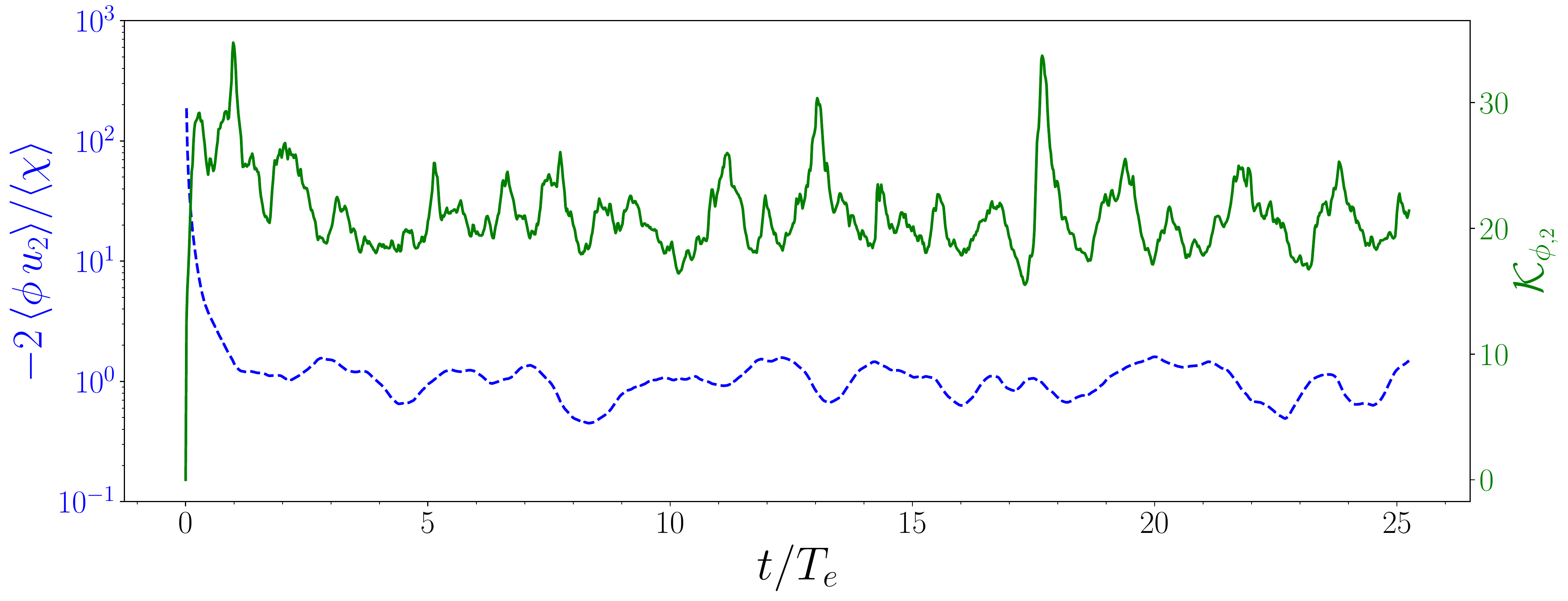}
        \subcaption{}\label{fig: scalar_snp}
    \end{minipage}
    \begin{minipage}[b]{.01\linewidth}
        ~
    \end{minipage}
    \begin{minipage}[b]{.34\linewidth}
        \centering
        \includegraphics[width=1\textwidth]{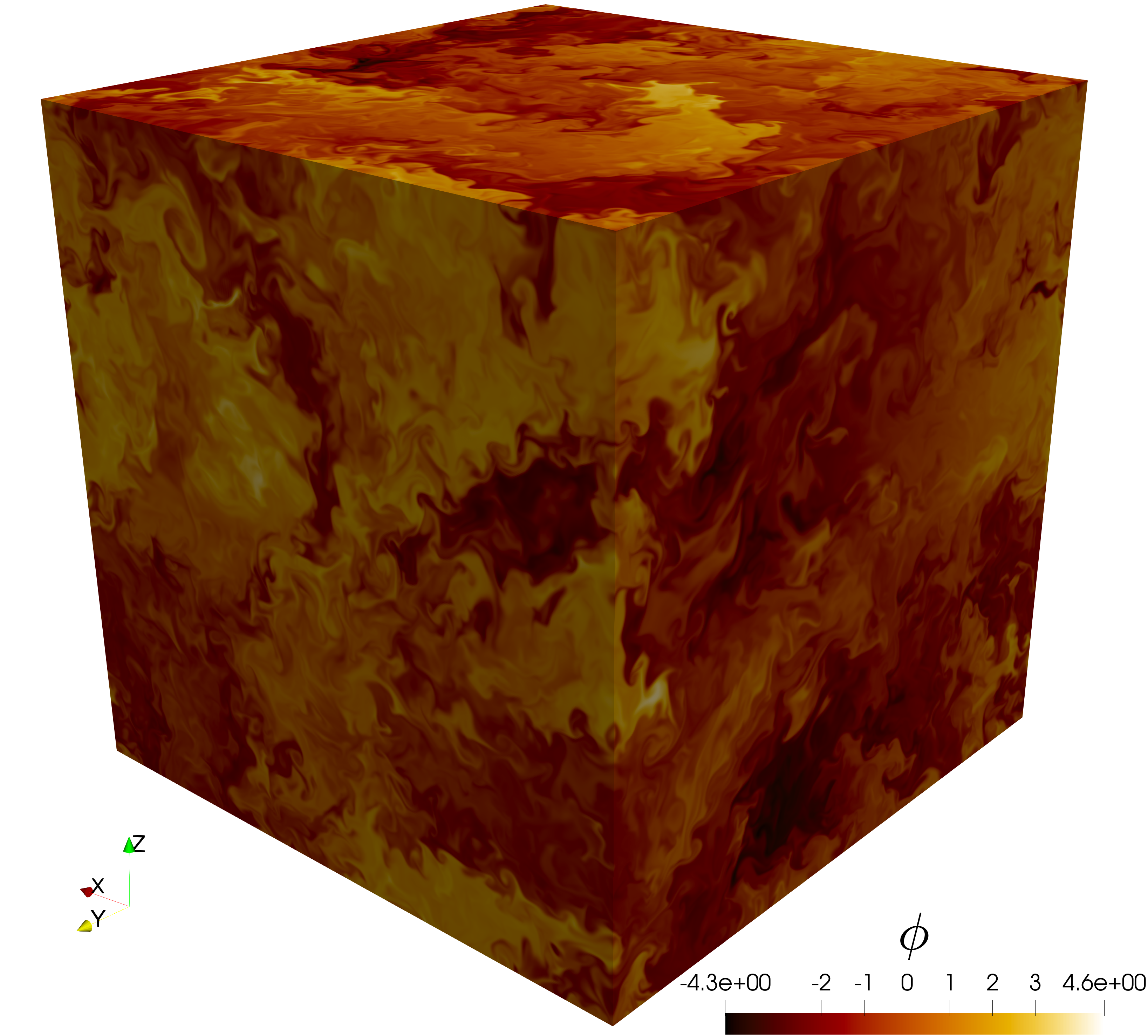}
        \subcaption{}\label{fig: Phi_snp}
    \end{minipage}
        % \centering
        % \includegraphics[width=.7\textwidth]{Scalar_Records}
    \caption{(a) Time-records of production over dissipation of scalar variance (blue dashed line), and the flatness factor for the scalar gradient vector component along the direction of mean scalar gradient (green solid line). (b) Snapshot of fully-developed turbulent passive scalar field.}\label{fig: Scalar_grad}
\end{figure*}

Since the fully-developed turbulent state is characterized by a meticulous tracking of statistical quantities of the flow, the present software provides a comprehensive framework for computing and recording the statistical quantities of turbulent flow. Given the homogeneity of the fluctuating fields, spatial averaging is employed for computing these records at user-defined time intervals. These statistical quantities are categorized into turbulent characteristics of small-scale motion reported in Table \ref{tab: small-scale_chr}, and high-order central moments of diagonal components of velocity gradient tensor (VGT), $\nabla \boldsymbol{u}$. For instance,
\begin{align}
    \label{eqn: Skw}
    \mathcal{S}_{u_{1,1}} =& \left. \left\langle \Big(\frac{\partial u_1}{\partial x_1}\Big)^3 \right\rangle \middle/ \left\langle \Big(\frac{\partial u_1}{\partial x_1}\Big)^2 \right\rangle^{3/2} \right.,
    \\ \label{eqn: Flt}
    \mathcal{K}_{u_{1,1}} =& \left. \left\langle \Big(\frac{\partial u_1}{\partial x_1}\Big)^4 \right\rangle \middle/ \left\langle \Big(\frac{\partial u_1}{\partial x_1}\Big)^2 \right\rangle^2 \right.,
\end{align}
where $\mathcal{S}_{u_{1,1}}$ and $\mathcal{K}_{u_{1,1}}$ indicate the skewness factor and flatness factor (or kurtosis) associated with the first diagonal component of VGT, $u_{1,1}=\partial u_1/\partial x_1$, respectively. Fully-turbulent flow state would be identified when the time-series of these records reach to a statistically stationary state after long enough time-integration, \textit{i.e.} approximately 10 to 15 large-eddy turnover times (see Table \ref{tab: small-scale_chr}). The parallel implementation for computing and collecting these statistical quantities and later recording them in time as different time series which were performed by \texttt{point-to-point} and \texttt{collective} MPI directives.

Furthermore, the velocity and pressure fields might be written as output files stored in directories named \texttt{Out}$\_\ast$ based on a user-defined time-interval that might be useful for any post-processing after the flow reaches to fully-developed turbulent state. A ``restart from file'' capability is also designated so that once the statistical record is written out on file, the latest state of velocity field and its related time-integration information are also output on files, which are stored in a directory named \texttt{Restart}. Starting a simulation from either a prescribed IC or restarting it to continue an ongoing simulation that was stopped is specified by a user-defined input argument. All the parallel I/O to store the velocity and pressure fields is done by employing \texttt{scipy.io} library and using \texttt{loadmat()} and \texttt{savemat()} routines for the partition of data that is resolved inside each MPI process in the compressed format and with the machine precision accuracy. This is a fast and efficient I/O while it maintains the simplicity for user for any post-processing step such as time-averaging on the statistically stationary turbulent data.

According to Figure \ref{fig: Schematic_Scalar}, once the fully-turbulent velocity state is achieved, the user might be able to use a restart or output instance as the velocity IC to introduce a passive scalar transport with a directional constant mean-gradient as described in \eqref{eqn: AD} while the fluctuating concentration is assumed to be zero, $\phi_0(\boldsymbol{x})=0$. Here, the goal is to resolve the fluctuating scalar concentration field, equation \eqref{eqn: P-S_AD}, transported on the fully-turbulent incompressible flow for long enough time-span so that the fully-developed and realistic turbulent state for the passive scalar is obtained. Subsequently, similar procedure as what described based on the schematic Figure \ref{fig: Schematic_HIT} is followed while the pseudo-spectral AD solver is fed by the resolved velocity field from the NS solver. The diffusivity of passive scalar, $\mathcal{D}$, is specified by a user-defined input argument for Schmidt number, $Sc=\nu/\mathcal{D}$. Accordingly, similar to the NS solver, the advective scalar flux, $(\reallywidehat{\boldsymbol{u}\cdot \nabla \phi})_{\boldsymbol{k}}$, is computed in the physical space by the inverse FFT of $\hat{\boldsymbol{u}}_{\boldsymbol{k}}$ and $(\reallywidehat{\nabla \phi})_{\boldsymbol{k}}$ and forward FFT computation of $\boldsymbol{u}\cdot \nabla \phi$. Similar dealiasing procedure as described for NS solver is employed in pseudo-spectral AD solver and the RK4 time-integration scheme is utilized to numerically perform explicit time-stepping.

In homogeneous scalar turbulence, time evolution of the scalar variance $\langle \phi^2 \rangle$ is governed by
\begin{align}\label{eqn: Scalar-var}
    \frac{d}{dt}\langle \phi^2 \rangle =& -2 \, \langle \boldsymbol{q} \rangle \cdot \nabla \langle \Phi \rangle - \langle \chi \rangle, 
\end{align}
where $\boldsymbol{q} = \phi \, \boldsymbol{u}$ denotes the scalar flux vector, and the scalar dissipation is defined as $\chi = 2 \, \mathcal{D} \, \nabla \phi \cdot \nabla \phi$ \citep{monin2013statistical}. According to \eqref{eqn: AD}, the first term in the right-hand side of \eqref{eqn: Scalar-var} is simplified to $-2 \, \beta \, \langle \phi \, u_2 \rangle$ that denotes the scalar variance production (by uniform mean scalar gradient, $\beta$). The present software is capable of computing and recording of rate of scalar variance in addition to the production and dissipation terms. This is useful in terms of the checking if the balance of both sides of equation \eqref{eqn: Scalar-var} holds throughout a simulation so that one ensures that the implementation of the solver works seamlessly. On the other hand, as a measure to evaluate that the statistically stationary state for the passive scalar is achieved is to check if $-2 \, \beta \, \langle \phi \, u_2 \rangle/\langle \chi \rangle \sim 1$ throughout the simulation. Moreover, recording the skewness and flatness factors for the components of fluctuating scalar gradient vector (\textit{e.g.}, $\mathcal{S}_{\phi_{,2}}$ and $\mathcal{K}_{\phi_{,2}}$ for $\phi_{,2}=\partial \phi/\partial x_2$ similar to \ref{eqn: Skw} and \ref{eqn: Flt} for VGT) is another statistical indicator measure for fully-developed turbulent passive scalar state. Therefore, in the current computational platform, the user would be able to recognize the statistically stationary state through monitoring the explained time-series data that is written out according to the user-defined time-interval as a software input.

The field data output and restart capability for the AD solver is designated similar to the described strategy for the NS solver so that the user would be able to resume an interrupted/stopped simulation and use the output field data for desired applications or post-processing.

In the following section, we present a comprehensive example that step-by-step walks through using the present software.

% =============================================================================
%                       Illustrative Examples
% =============================================================================
\section{Illustrative Example}\label{sec: Examples}

This comprehensive example is mainly consisted of construction of isotropic velocity IC, obtaining well-resolved fully-turbulent velocity field, and simulating well-resolved passive scalar turbulence with imposed mean scalar gradient.

\subsection{IC construction and DNS of the HIT}\label{sec: ex-HIT}

\begin{table}[t!]
	\centering
	    \caption{Input arguments for the \texttt{PScHIT.py} and the specified values for the example case. The order of the arguments in the execution command-line are as listed here.}\label{tab: inp_args}
	    \centering
    	\begin{tabular}{c c c}
    		\toprule \toprule
    		Input Argument & $\qquad$ & Value \\ \midrule
    		\texttt{t\_end}& $\qquad$ &  40\\
    		\texttt{Output\_frequency}& $\qquad$ &  1000\\
    		\texttt{Stats\_frequency}& $\qquad$ &  100\\
    		\texttt{TKE\_magnification}& $\qquad$ &  6.0\\
    		$\nu$& $\qquad$ &  0.0008\\
    		$k_F$& $\qquad$ &  2\\
    		\texttt{forcing\_type} & $\qquad$ & deterministic\\
    		$N$& $\qquad$ &  520\\
    		$\Delta t$& $\qquad$ &  0.0005\\
    		\texttt{If\_Restart} & $\qquad$ &  0 or 1\\
    		\bottomrule \bottomrule
    	\end{tabular}
\end{table}

According to the descriptions in section \ref{sec: Architecture}, the isotropic and divergence-free velocity IC is constructed based upon a prescribed energy spectrum given in \eqref{eqn: TKE_Spectrum_IC}. Considering the periodicity length $\mathcal{L}=2\pi$, this pre-processing step is done through serial execution of \texttt{Gen\_IC.py} script that takes the following input arguments, respectively: $N$ (spatial resolution along each direction), $k_F$ (forcing wavenumber), and $N_p$ (number of slab partitions). We need to emphasize that $N_p$ must be chosen in a way that $N$ be a multiple of $N_p$. The resulting velocity field is located in a directory named \texttt{IC}, where $N_p$ number of $\texttt{.mat}$ velocity files are stored. In this example, we take $N=520$, $k_F=2$, and $N_p=40$. All the components of velocity IC in addition to the VGT components have Gaussian distribution. This velocity IC is being passed into the NS solver written in \texttt{PScHIT.py} script that takes the following input arguments given in Table \ref{tab: inp_args}. Here, \texttt{Output\_frequency} and \texttt{Stats\_frequency} are multiplied by the specified $\Delta t$. Moreover, \texttt{If\_Restart} argument could be either 0 or 1, where 0 indicates it is a simulation starting from the constructed IC while 1 specifies resuming a simulation from restart files. In this example, we perform the simulation for $t/T_e \sim 15$ to ensure the fully-turbulent flow state is achieved and Figure \ref{fig: HIT_u1} portrays the first component of the velocity field. Figure \ref{fig: TKE_spectrum} shows the radial TKE spectrum averaged over 5 large-eddy turnover times. Moreover, Figure \ref{fig: HIT_records} includes the time-records of the Taylor-scale Reynolds number and VGT skewness factor for diagonal components computed and recorded over 40 large-eddy turnover times. This shows that the statistically stationary state is achieved through the long-time DNS where $Re_\lambda \sim 240$, $\mathcal{S}_{u_{1,1}}=\mathcal{S}_{u_{2,2}}=\mathcal{S}_{u_{3,3}} \sim -0.55$, and $\mathcal{K}_{u_{1,1}}=\mathcal{K}_{u_{2,2}}=\mathcal{K}_{u_{3,3}} \sim 6.8$ at fully-turbulent state. The statistical records of VGT clearly show that the resolved velocity field is isotropic. Finally, $k_{max}\eta > 1.45$ ensures that the small scale turbulent motions are well-resolved ($k_{max}=\sqrt{2}N/3$) \citep{Overholt1996}. 

\subsection{DNS of passive scalar transport}\label{sec: ex-Scalar}

Similar to starting the NS solver from a prescribed velocity IC, we take a fully-turbulent velocity output (velocity state at $t/T_e=15$ in section \ref{sec: ex-HIT}) and continue the simulation under the artificial forcing while we introduce a passive scalar field where its fluctuating part is initialized at zero. The Schmidt number, $Sc$, is specified by user through an input argument. According to the problem setting for the mean scalar gradient, we let $\beta = 1$ (mean scalar gradient along $x_2$ direction). Therefore, for a passive scalar with $Sc=1$, we aim to obtain the fully-turbulent scalar field. We need to note that the spatial resolution required for the passive scalars with $Sc \geq 1$ is defined based on $\eta_B=\eta \, Sc^{-1/2}$ \citep{batchelor1959small} and in this example the spatial resolution for the velocity field is sufficient for a well-resolved passive scalar. We manage to resolve the passive scalar field for 25 large-eddy turnover times and the rest of the simulation parameters remain the same as values reported in the Table \ref{tab: inp_args}. Figure \ref{fig: Scalar_grad} shows the records of scalar variance production over dissipation rate, $-2 \, \langle \phi \, u_2 \rangle/\langle \chi \rangle$, and the flatness factor for the scalar gradient along the direction of mean scalar gradient, $\mathcal{K}_{\phi_{,2}}$. As it is observed, after resolving the passive scalar field for approximately two large-eddy turnover times, $-2 \, \langle \phi \, u_2 \rangle/\langle \chi \rangle \sim 1.0$ that means the equilibrium state for the passive scalar variance is obtained. Moreover, after approximately three large-eddy turnover times the high-order statistical moments of the scalar gradient reach to a statically stationary state. For instance, $\mathcal{S}_{\phi_{,2}}\sim 1.4$, and $\mathcal{K}_{\phi_{,2}}\sim 20.8$ throughout the time-averaging of these statistical moments when $t/T_e \geq 5$. By resolving the passive scalar field through AD  equation and for long enough time after the equilibrium and stationary state, the fully-turbulent and realistic scalar field is ensured.

% =============================================================================
%                           Impact
% =============================================================================
\section{Impact}\label{sec: Impact}

Current work offers a framework to obtain highly accurate spatio-temporal data for homogeneous turbulent transport with proper statistical testing from the recorded quantities. In turbulent transport research, this provides a great source of high-fidelity data for a variety of innovative contributions. For instance in large-eddy simulation (LES) of turbulent transport, novel nonlocal models for the subgrid-scale (SGS) stress and flux terms appearing in the filtered NS and AD equations heavily depend on the DNS data for such transport phenomena to compute the exact values of SGS terms to evaluate the model performance \citep{samiee2020fractional}. On the other hand, in the abundance of data and emergence of the data-driven turbulence models \citep{duraisamy2019turbulence}, current computational platform would be a reliable candidate to generate data for training and testing such models \citep{wang2018investigations, beck2019deep, portwood2020interpreting, Mohan_JoT2020, mohan2020embedding, beetham2020formulating, SIRIGNANO2020}. Moreover, high-Reynolds and well-examined high-fidelity turbulent transport data from the present platform could be directly employed in studying the role of coherent turbulent structures and their effects on the turbulence statistics \citep{zayernouri2011coherent, akhavan2020anomalous}, as well as investigating topological characteristics of turbulent transport \citep{gonzalez2009kinematic, meneveau2011lagrangian}, analysis of extreme events and the internal intermittency \citep{yeung2015extreme, farazmand2017variational, sapsis2020statistics}, and diverse modeling strategies for the arising closures \citep{das2019reynolds, das2020revisiting}.

% =============================================================================
%                           Conclusions
% =============================================================================
\section{Conclusion}\label{sec: Conclusion}

This work presented a computational platform for DNS of homogeneous turbulent flow and passive scalar transport. This open-source software works based upon a pseudo-spectral representation of the NS and AD equations on a triply cubic computational domain with periodic boundary conditions for fluctuating fields. Using Fourier collocation method, the governing equations are discretized in space and by employing RK4 scheme the time-stepping is performed. The software provides a pre-processing step to construct homogeneous and isotropic divergence-free velocity IC based on prescribed energy spectrum and decompose it into user-defined partitions. Using artificial forcing scheme, the dissipated energy is injected to the low wavenumbers so that after long-time integration, the statistically stationary state is achieved. In order to examine and identify if the fully-developed turbulent flow is obtained, small-scale statistical quantities of turbulence in addition to the central moments of VGT components are computed and recorded. Once the realistic turbulent velocity field is obtained, the user is able to start resolving a passive scalar that is transported on the HIT flow while a uniform mean scalar gradient is imposed. Resolving the scalar fluctuations for long enough time after reaching to the equilibrium and stationary state provides the fully-developed turbulent scalar field. Statistical records of scalar gradients in addition to the records of production and dissipation of scalar variance helps the user to properly identify when the fully-developed scalar turbulence is achieved.

% =============================================================================
%                       Acknowledgement
% =============================================================================
\section*{Acknowledgement}
This work was financially supported by the MURI/ARO award (W911NF-15-1-0562), the ARO Young Investigator Program (YIP) award (W911NF-19-1-0444), and the NSF award (DMS-1923201). The HPC resources and services were provided by the Institute for Cyber-Enabled Research (ICER) at Michigan State University.

% =============================================================================
% =============================================================================

%\bibliographystyle{elsarticle-num-names}
\bibliographystyle{elsarticle-num}
\bibliography{mybibfile}

\end{document}